\documentclass[prb,twocolumn]{revtex4-2}
\usepackage[utf8]{inputenc}
\usepackage[T1]{fontenc} 
\usepackage{graphicx} 
\usepackage[version=4]{mhchem}
\usepackage{dcolumn}
\usepackage{booktabs, array, mathptmx, float, tabularx, booktabs, lipsum, amsmath,multirow}
\usepackage{siunitx, xcolor}

\begin{document}

\preprint{APS/123-QED}

\title{Superatom Orbitals, Orbital Splitting and Structure Prediction of Pure Alkali Metal Clusters}
\author{Jin Liu , Zhi-Jie Yan , YI-Chao Jin }
\affiliation{Department of Science, East China University of Science and Technology, Shanghai}
\author{Meng Zhang}
\email{mzhang@ecust.edu.cn}
\affiliation{Department of Science, East China University of Science and Technology, Shanghai}
\date{\today}
\begin{abstract}
Jellium model achieved great success in predicting stable clusters with closed electronic shells and zero spin. In order to explain the stability of open shell clusters, it is necessary to consider the case of non-degenerate energy levels. In this paper the energy levels in nine low-lying Li$_{19}$ clusters are analysed systematically through superatomic orbital splitting effect. It is found that for originally degenerate orbitals like five 1D orbitals, the more the orbital extends in the direction of the cluster extension, the lower the energy of the orbital becomes.  So oblate Li$_{19}$ clusters have the orbital sequence of $1\mathrm{S}^2-1\mathrm{P}_{x/y}^{4}-1\mathrm{P}_{z}^{2}-1\mathrm{D}_{xy/x^2-y^2}^{4}-2\mathrm{S}^2-1\mathrm{D}_{xz/yz}^{4}-1\mathrm{D}_{z^2}^{1}$, while prolate Li$_{19}$ clusters have the sequence of  $1\mathrm{S}^2-1\mathrm{P}_{z}^{2}-1\mathrm{P}_{x/y}^{4}-1\mathrm{D}_{z^2}^{2}-1\mathrm{D}_{xz/yz}^{4}-1\mathrm{D}_{xy/x^2-y^2}^{4}-2\mathrm{S}^1$. This electron configuration is applied to predict the shape and magnetic moment of the alkali metal Li$_{n}$ clusters. The stability of the Li$_{14}$ cluster can be successfully interpreted in the framework of orbital splitting effect without resorting to the super valence bond (SVB) model, indicating a non-spherical cluster can achieve good stability without meeting the magic number. It is also proposed that the orbital splitting can be used to predict the shape (prolate, oblate or sphere) and magnetic moment of clusters. 11 out of 16 predicted shapes of Li$_n(n=3-18)$ are consistent with the results obtained by the principle of minimum energy.
\end{abstract}
\maketitle

\section{INTRODUCTION}
Jellium model assumes that the movement of valence electrons in simple metal clusters is carried out in a uniform potential\cite{ekardt_work_1984, ekardt_theory_1986}. The quantum states of valence electrons in the cluster follows the configuration of $1\mathrm{S}^2-1\mathrm{P}^6-1\mathrm{D}^{10}-2\mathrm{S}^2-1\mathrm{F}^{14}-2\mathrm{P}^6-\cdots $, with S, P, D, F the angular momentum. This way, these clusters, also known as superatoms, mimic the chemical behaviour of single atom \cite{shao_density_2015, yan_face-sharing_2019, yan_superatomic_2020, sung_packing_1994, zhang_magnetic_2013, zhang_probing_2010}. Jellium model predicts that superatoms with full electronic shells are endowed with extra stability, giving the magic numbers of 2, 8, 18, 20, 34, $\cdots$. For example, Li$_3$O$^+$ cluster consists of 8 electrons, responsible for a gap as high as 3.72 eV between the highest occupied molecular orbital (HOMO) and the lowest unoccupied molecular orbital (LUMO) \cite{pauna_evolution_2017}. Tetrahedral Au$_{20}$ cluster is of special stability, of which the electronic shell can be viewed as $1\mathrm{S}^2-1\mathrm{P}^6-2\mathrm{S}^2-1\mathrm{D}^{10}$ \cite{cheng_2014_superatom}. Experimental measured HOMO-LUMO gap is 1.77 eV, greater than that in C$_{60}$ \cite{li_au20_2003}. 40-electron Al$_{13}^-$  cluster is known for mimicing hologen atoms, the stability of which originates from closed 2P shell\cite{ bergeron_formation_2004}. \\
Despite the great success that shperical jellium model has achieved on the explanation of stability possessed by closed shell superatoms, the limitation of magic number hinders the diversity of physical and chemical properties of clusters\cite{reveles_2009_designer}. Here we propose that when orbitals are splitted into several subgroups, extra stability can also be achieved if electrons happen to fully or halfly occupy one subgroup. In this work, we report the splitting of originally degenerate superatom orbitals through our investigation of orbitals in Li$_{19}$ clusters, the number of electrons $n=19$ is  exactly between  magic numbers 18 and 20. When a transition metal cation is located in the center of an octahedral cage of oxygen anions, due to the electric repulsion produced by ligands, its d orbitals splits into two sets, which is the core idea of crystal field theory \cite{kugel_1982_jahn}. Orbitals in superatoms, e.g., 1P and 1D, will also split into several groups, only the intrinsic dirving force is not ligand's electric field but the deformation of cluster shape.  \\
Lithium, the simplest alkali metal with electronic shell of $1\mathrm{s}^2-2\mathrm{s}^1$, is an ideal prototype for simple metal \cite{cheng_2013_communication, yan_seventeen-coordinate_2019, yan_face-sharing_2019, yan_superatomic_2020}. Li$_{19}$ cluster owns 19 electrons, and its electrons can fully occupy orbitals from 1S to 1D, providing convenience for analysis of electronic orbitals. Based on a combined searching strategy and DFT calculation, the optimal configuration of Li$_{19}$ cluster is obtained along with eight low-lying isomers. The superatom orbitals are studied and it is found that their electronic configurations can be divided into two types, corresponding to two different cluster shapes, prolate ellipsoid and oblate ellipsoid, respectively. Clusters in the shape of oblate ellipsoid have the electronic configuration of $1\mathrm{S}^2-1\mathrm{P}^6-1\mathrm{D}_{xy/x^2-y^2}^4-2\mathrm{S}^2-1\mathrm{D}_{xz/yz}^4-1\mathrm{D}_{z^2}^1$, while clusters in the shape of prolate ellipsoid have the configuration of  $1\mathrm{S}^2-1\mathrm{P}^6-1\mathrm{D}_{z^2}^2-1\mathrm{D}_{xz/yz}^4-1\mathrm{D}_{xy/x^2-y^2}^4-2\mathrm{S}^1$. We conclude that for originally degenerate orbitals, the more the orbital extends in the direction of the cluster extension, the lower the energy becomes. It is found that this relationship can be applied to explain the stability of Li$_{14}$ and predict the shape as well as magnetic moment of alkali clusters. Our results based on superatomic orbital splitting effect analysis is a powerful tool and theoretically sound for explaining and predicting the structure and stability of simple alkali metal clusters.

\section{COMPUTATIONAL DETAILS}
A large number of initial isomers are generated by the following methods in order to carry out an unbiased search for the global-minimum structures of Li$_{19}$: \\
\begin{enumerate}
\item Crystal structure AnaLYsis by Particle Swarm Optimization (CALYPSO) software package developed by Yanchao Wang et.al \cite{lv_2012_particle}.
\item Blending smaller clusters. Place several smaller clusters (19 lithium atoms in total) at random relative position, which simulates the formation of large clusters through collision of small clusters. 
\item Constructing geometrically symmetric system. We randomly fabricated polygons with high mathematical symmetry and fit lithium atoms in geometric vertex properly. 
\end{enumerate}
All Electron Relativistic methods were subsequently performed for further geometric optimization with Dmol3 package in $Materials Studio$. The Perdew and Wang’s 1991 exchange and correlation functional (PW91) functional \cite{wang_1991_correlation} was continually selected since the quality has been verified by our previous work in V@Li$_n$ superatom clusters \cite{zhang_magnetic_2013}. The highest precision basis set of double numerical plus polarization with addition of diffuse functions (DNP+) was chosen. Cut-off energy and the self-consistent field (SCF) convergence tolerance were both set as Fine. The frequencies of the configurations were calculated and there are no imaginary frequencies for the obtained configuration with lowest energy. For odd or even number of valence electrons, the potential spin multiplicity may be 2, 4, 6, 8... or 1, 3, 5, 7..., respectively. Each possible spin multiplicity of the clusters was tested and value with the lowest energy is picked. \\

\section{RESULTS AND DISCUSSION}
\subsection{Identification of Geometric Configurations}
\begin{figure}[htbp]
\centering
\includegraphics[scale=0.07]{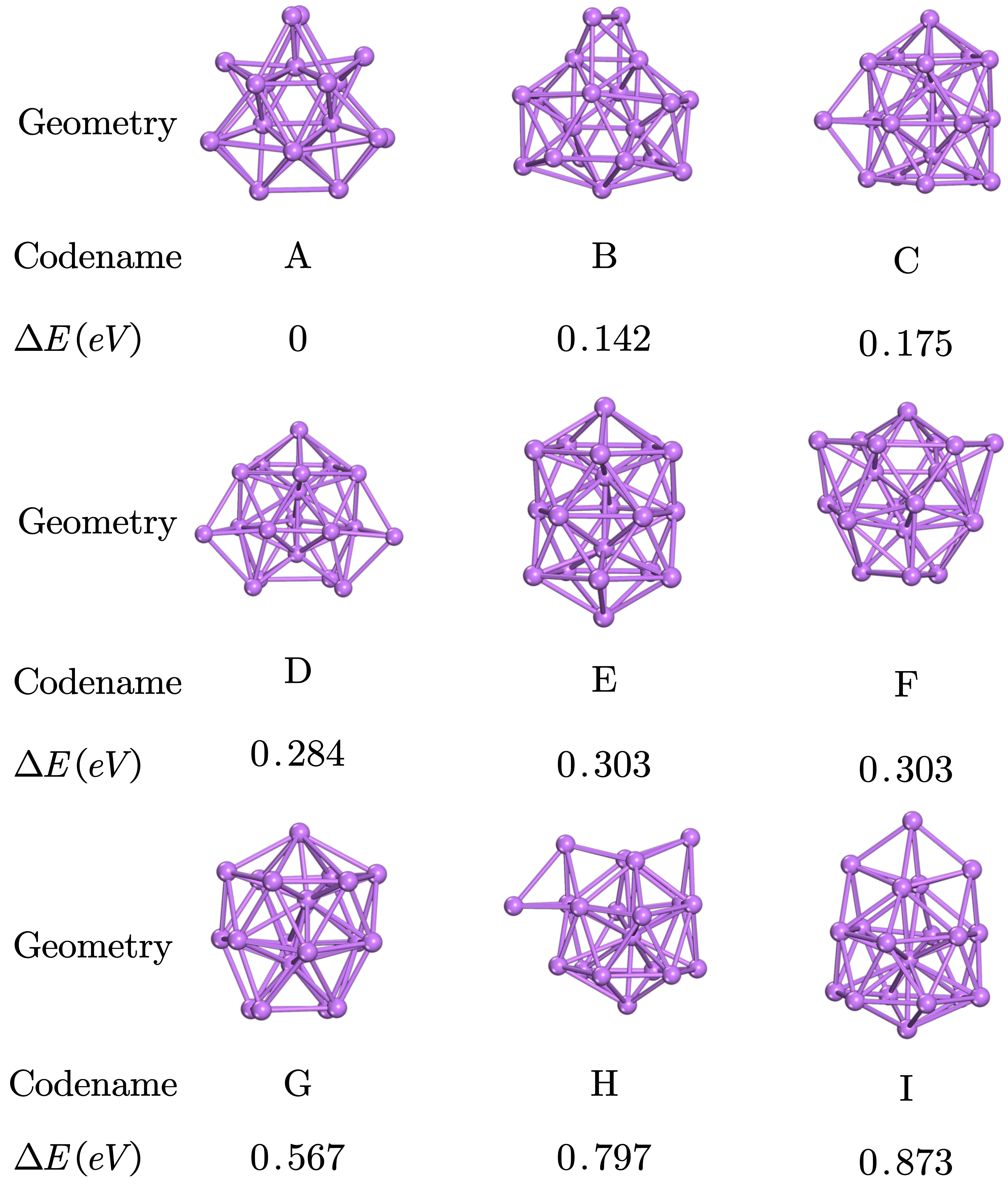}
\caption{Nine low energy structures of Li$_{19}$ marked by letter A to I. Their relative energies are calculated.}
\label{Li19_stru}
\end{figure}

Although a large variety initial structures were obtained from the above three different methods, the lowest energy structrue obtained by further DFT optimization of these initial configurations are indeed the same. We listed 9 isomers with distinct geometry in FIG. \ref{Li19_stru} sorted by energy from low to high. \\
The ground state structure for Li$_{19}$ clusters is an oblate cage with three lithium atoms inside as marked A in FIG. \ref{Li19_stru}, which is consistent with the result predicted by Sung using simulated annealing method \cite{sung_packing_1994}. According to the jellium model, alkali metal clusters with closed-shell electronic configurations usually adopt high symmetry geometries \cite{luo_special_2014}. Since Li$_{19}$ cluster has total 19 valence electrons, which is between two magic number of 18 and 20, it is in line with expectations that structure of the Li19 has low C$_{2v}$ symmetry and does not tend to form a spherical shape. Another intriguing structure is E with the symmetry of  D$_{5h}$. Its shape is a prolate cage-like structure with a morphology similar to that of Chinese lanterns. Four atoms sit on its central axis, encircled by three staggered 5-membered rings. C and H structures evolve from E and I structures respectively with one atom moves from the top or buttom of the original structures to their side. These two isomers C and H are considered as the transitiaoal structures because they are  because it is somewhere in between prolate ellipsoid and oblate ellipsoid. Binding energies of  oblate C and H are lower than their parent structures E and I, respectively, indicating that Li$_{19}$ is more inclined to form oblate ellipsoid clusters. This structural characteristic is also verifed by the most stable isomer A with the oblate structure. \\
\subsection{Superatom Orbitals in Li$_{19}$}
\begin{figure*}
\centering
\includegraphics[scale=0.1]{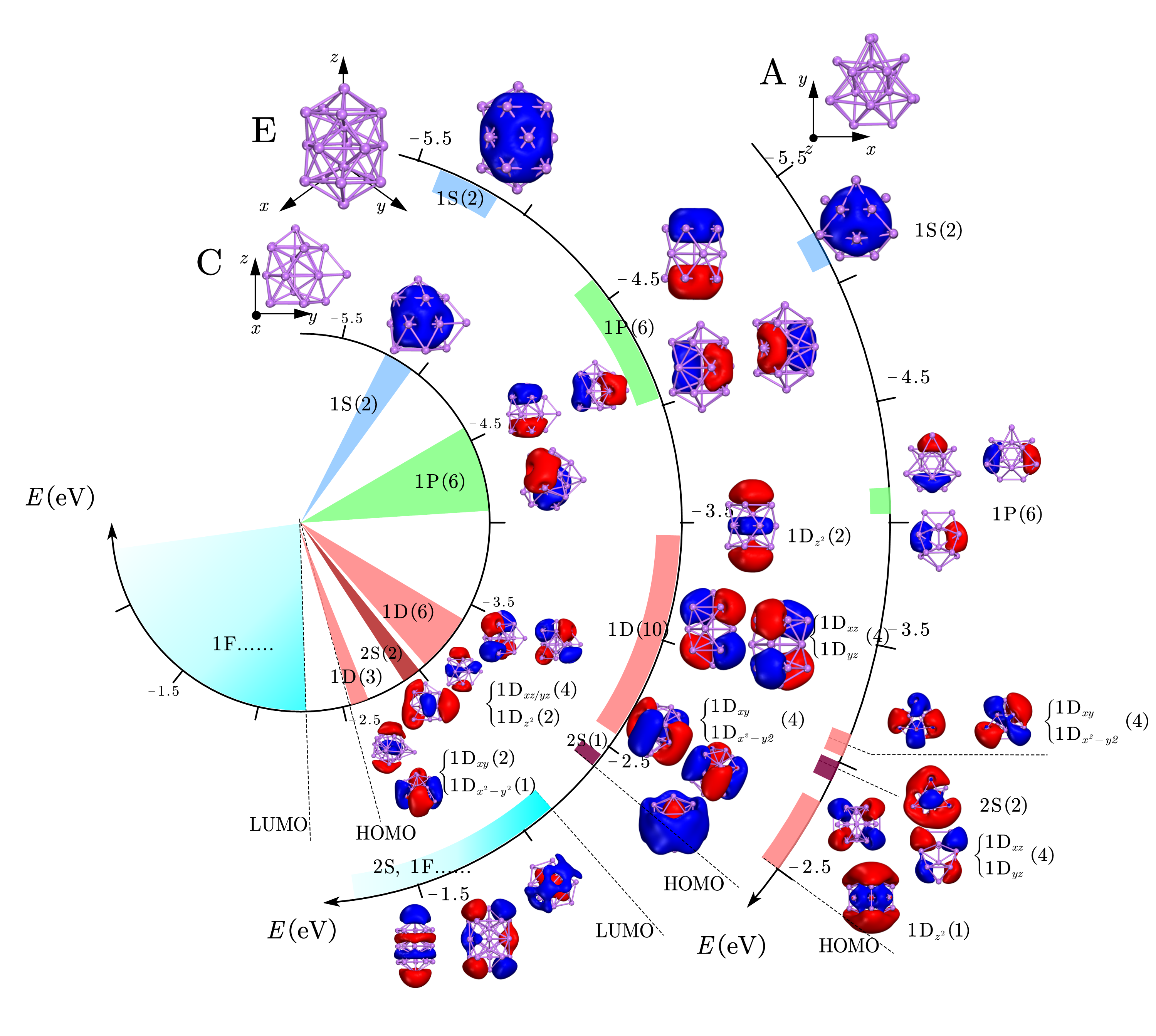}
\caption{The molecular orbitals (MOs) and MO energy level diagrams of the pristine Li19 cluster for structure C, E and A respectively. The color bar on the scale axis represents the energy range covered by corresponding sets of MOs. The shapes of MOs are drawn outside the scale axis, and the type and number of each MOs are marked nearby. To represent the energy level order for non-degenerate orbitals, the orbitals drawn before are lower in energy while the orbitals in the same radial position have similar energies. The highest occupied molecular orbital (HOMO) and lowest unoccupied molecular orbital (LUMO) energy levels are marked by dotted lines.}
\label{orbital2}
\end{figure*}
\begin{figure*}
\centering
\includegraphics[scale=0.6]{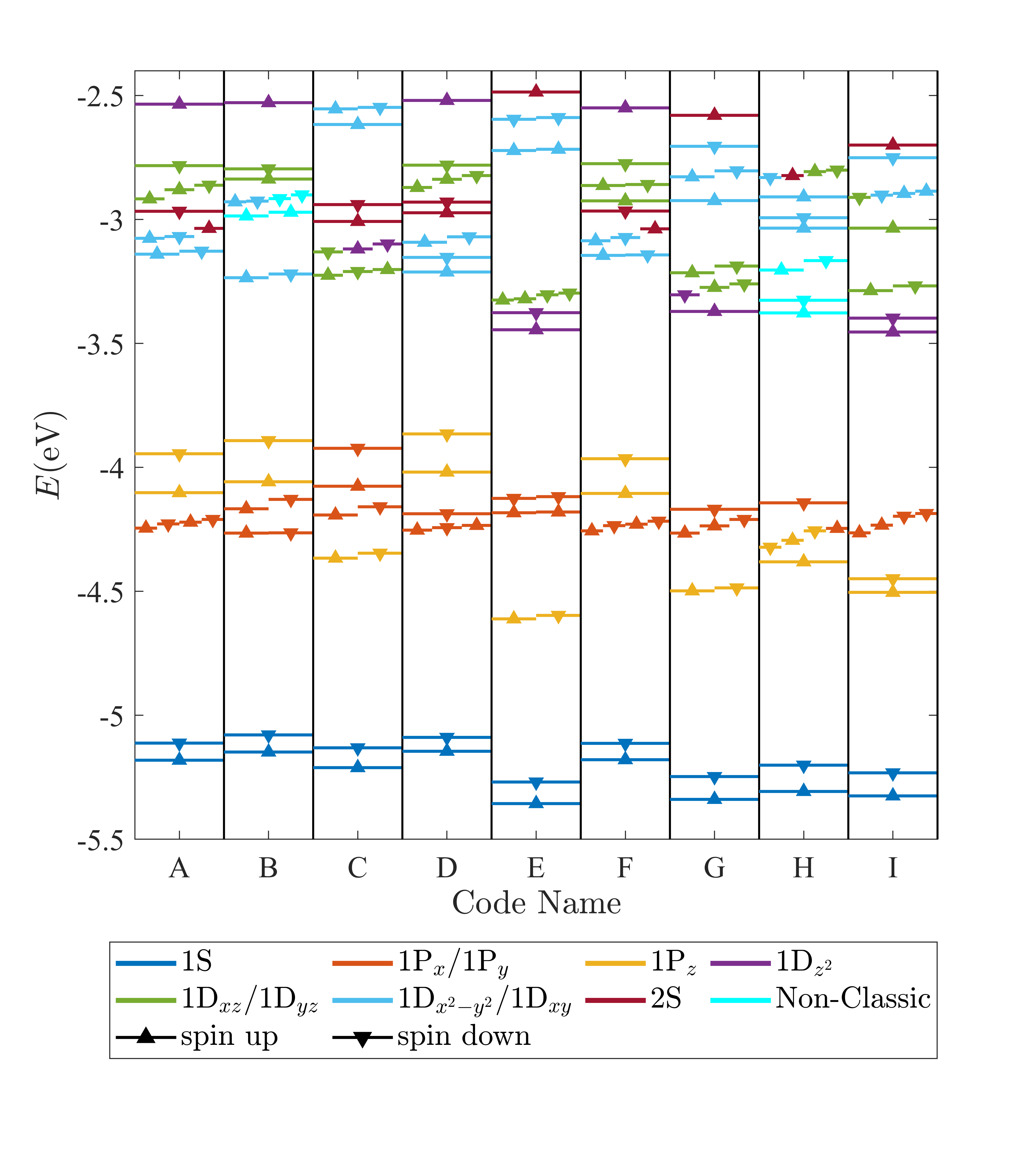}
\caption{Energy level of superatom orbitals of Li$_{19}$ clusters where solid lines with different colors represent different energy levels. Orbitals with strange shapes that cannot be classified as known orbitals are classified as Non-Classical.}
\label{Li19level}
\end{figure*}
Orbital analysis reveals that Li$_{19}$ clusters possess superatom orbitals and mimic the s, p, d orbitals of simple atom. As each lithium atom provides one delocalized electron, the number of effective valence electrons is 19. Despite having a open shell, it is found that molecular orbitals of Li$_{19}$ clusters can be discussed under jellium model. FIG. \ref{orbital2} displays orbitals of three typical structures: A, C and E as representatives of oblate cluster, transition shape cluster and prolate cluster respectively. All these three sets of orbitals show a lot of similarities to the atomic orbitals.The spherical 1S orbital is followed by the dumbbell-shaped 1P orbital, then comes the petal-like 1D orbital together with 2S orbital. The identification of orbitals relies on the orientation of the coordinate axes, which is determined by the following method: we take advantage of co-planar 1D$_{xy}$/1D$_{x^2-y^2}$ to fix x-y plane, thus the direction penperdicular to the plane is z axis. \\
As one might note the orbital alignment sequences of three isomers in FIG. \ref{orbital2} differ from one another. For the most stable configuration of pure Li$_{19}$ marked as A, 1D orbitals split into three groups, $1\mathrm{D}_{xy}/1\mathrm{D}_{x^2-y^2}$ with the lowest energy, $1\mathrm{D}_{xz}/1\mathrm{D}_{yz}$ with the second lowest energy, and $1\mathrm{D}_{z^2}$ with the highest energy. 1P orbitals also splitted into two groups and the energy of $1\mathrm{P}_{x/y}^{}$ is lower than the energy of $1\mathrm{P}_z$. The more the orbital stretches in the $z$ direction, the higher the energy become. For prolate configuration E, the sequence of splitted orbials is dramatically opposite. 1D orbitals with lowest energy are $1\mathrm{D}_{z^2}$, followed by $1\mathrm{D}_{xz}/1\mathrm{D}_{yz}$, and highest energy level orbitals are $1\mathrm{D}_{xy}/1\mathrm{D}_{x^2-y^2}$. The situation is same in 1P orbitals. The more the orbital stretches in the z direction, the lower the energy. Orbitals in transition structure C basically follows the pattern of of the prolate ellipsoid, but also begins to show some signs of transition. For example, the energy of $1\mathrm{D}_{z^2}$ is higher than that of $1\mathrm{D}_{xz}/1\mathrm{D}_{yz}$ in isomer C. Is this a random case or is there an underlying pattern? To explore the underlying pattern of superatom orbitals, the detailed molecular orbitals and orbital energies of the clusters are calculated and identified. FIG. \ref{Li19level} gives a general overview of energy levels of nine low energy structures of Li$_{19}$ given in FIG. \ref{Li19_stru}. It shows that the energy levels are nondegenerate and split in certain order. For geometry A, B, D and F, the orbital sequence is $1\mathrm{S}^2-1\mathrm{P}^6-1\mathrm{D}_{xy/x^2-y^2}^4-2\mathrm{S}^2-1\mathrm{D}_{xz/yz}^4-1\mathrm{D}_{z^2}^1$. For geometry E, G and I, the sequence is $1\mathrm{S}^2-1\mathrm{P}^6-1\mathrm{D}_{z^2}^2-1\mathrm{D}_{xz/yz}^4-1\mathrm{D}_{xy/x^2-y^2}^4-2\mathrm{S}^1$. For C, as analysed previously, the orbitals mainly draws out the properties of prolate clusters along with the transition. For H, because of the irregular geometric structure, the order of its energy levels is also not obvious. Besides, the exsistance of nonclassical orbitals caused by  irregular geometry also hindered the summarization of the pattern. \\
\begin{figure*}
\centering
\includegraphics[scale=0.15]{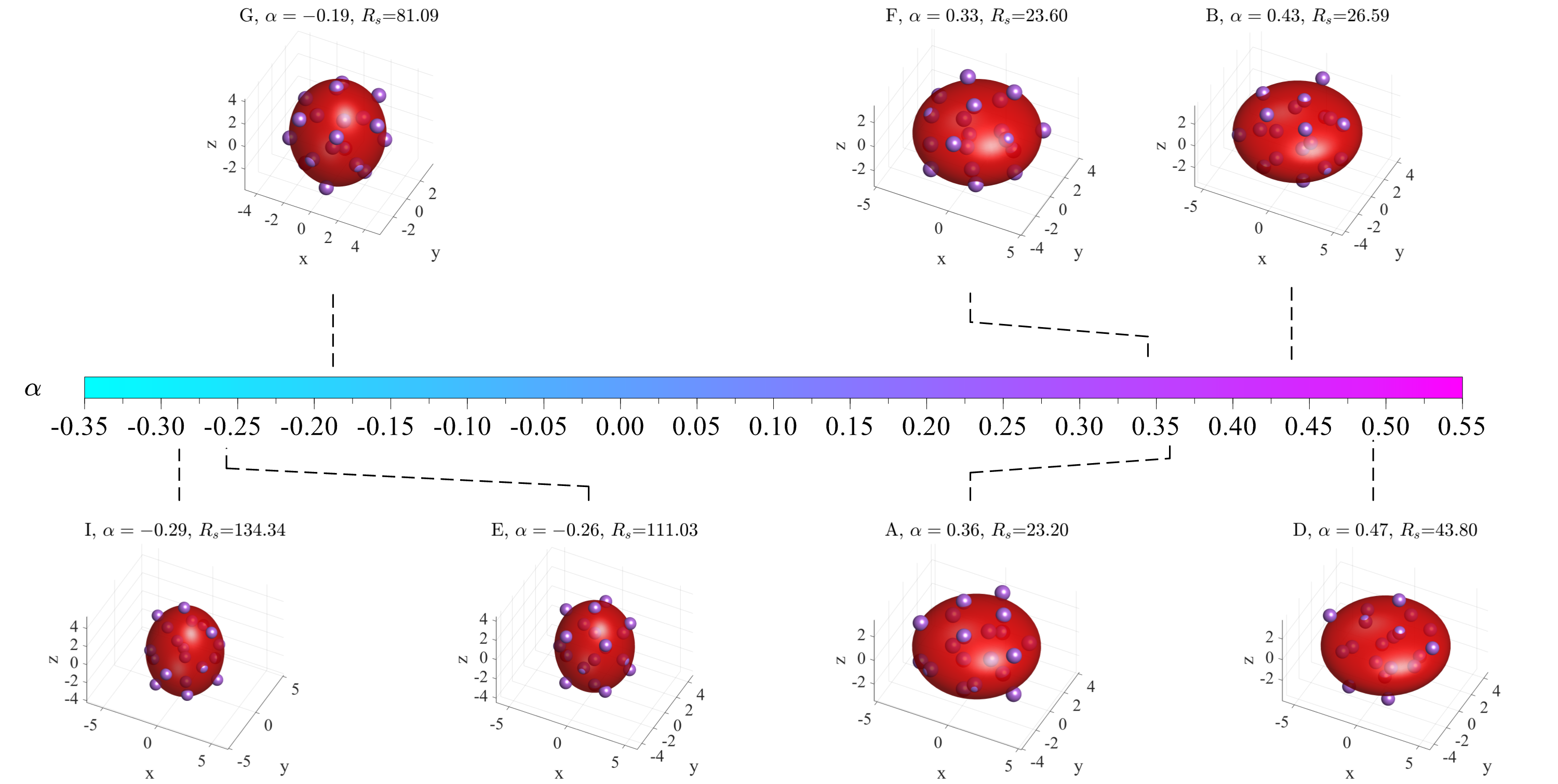}
\caption{Fitting results of the flattening of seven lithium clusters, atoms inside the cluster have a weight of 0 in the fit. $\alpha =\left( a-b \right) /a$ describes the flattening of the ellipsoid and $R_s$ is the squared norm of the residual. }
\label{alpha}
\end{figure*}
We linked the pattern to the flattening of ellipsoid of cluster shapes. Function $\left( x^2+y^2 \right) /a^2+z^2/b^2=1$ is applied to fit the shape of cluster as an rotational ellipsoid and the flattening is defined as $\alpha =\left( a-b \right) /a$ as Clemenger pointed out in 1985 \cite{clemenger_ellipsoidal_1985}. For prolate ellipsoid $\alpha <0$ while for oblate ellipsoid $\alpha > 0$. As in FIG. \ref{alpha}, geometry A, B, D, F are oblate ellipsoids with $\alpha$ ranging from 0.33 to 0.47, geometry E, G, I are prolate ellipsoids with $\alpha$ ranging from -0.29 to -0.19. By such classification, we can draw the conclusion that oblate clusters have the orbital sequence of $1\mathrm{S}^2-1\mathrm{P}^6-1\mathrm{D}_{xy/x^2-y^2}^4-2\mathrm{S}^2-1\mathrm{D}_{xz/yz}^4-1\mathrm{D}_{z^2}^1$, while prolate clusters have the sequence of  $1\mathrm{S}^2-1\mathrm{P}^6-1\mathrm{D}_{z^2}^2-1\mathrm{D}_{xz/yz}^4-1\mathrm{D}_{xy/x^2-y^2}^4-2\mathrm{S}^1$. For originally degenerate orbitals, the more the orbital extends in the direction of the cluster extension, the lower the energy becomes (e.g. energy of 1D$_{z^2}$ is lower than 1D$_{xz/yz}$ in prolate cluster because 1D$_{z^2}$ has a greater lob in $z$ axis). \\
\subsection{Prediction of Stability}
\begin{figure}[H]
\centering
\includegraphics[scale=0.08]{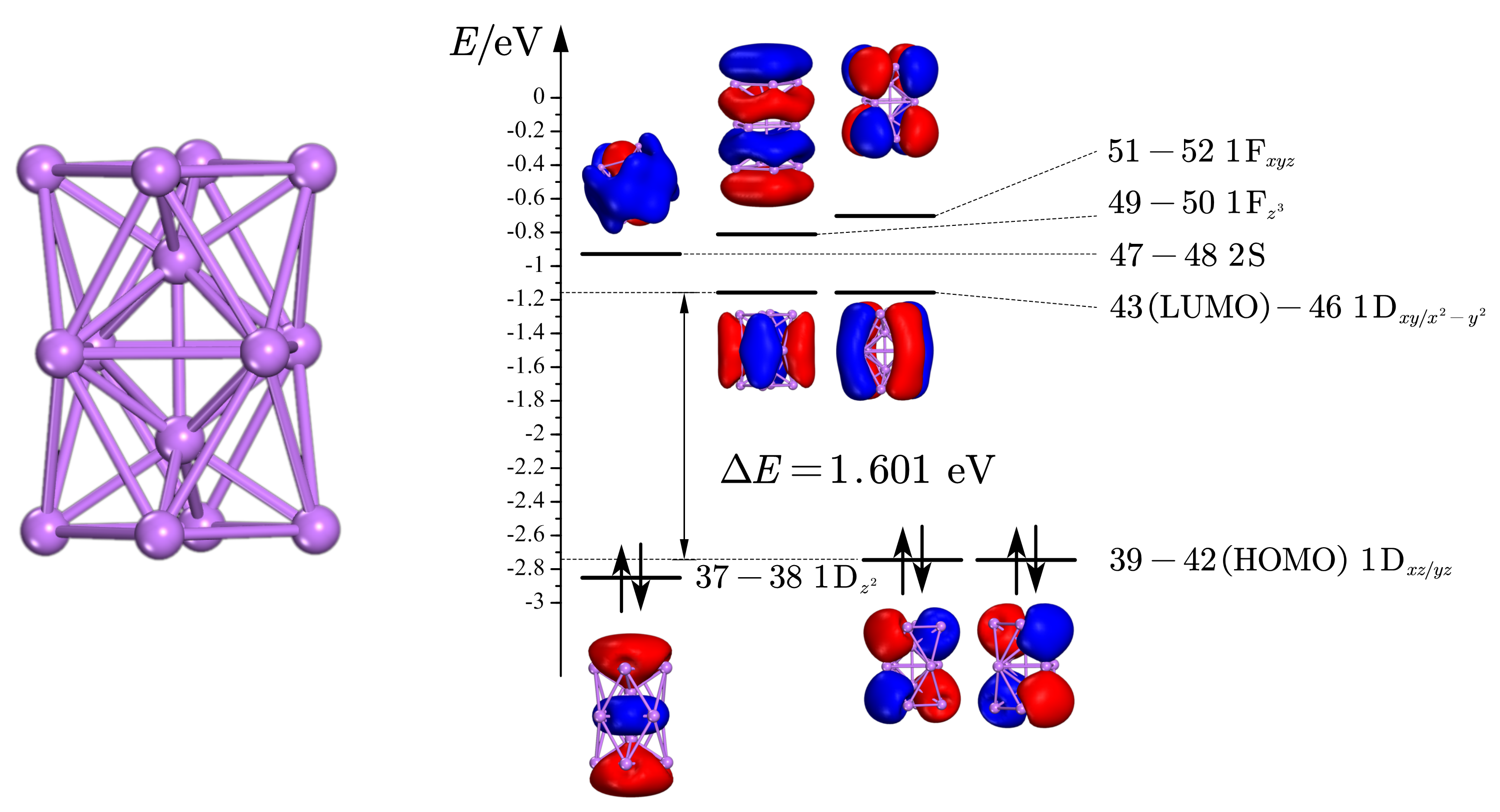}
\caption{Calculated energy levels and electronic orbitals in Li$_{14}$ clusters.}
\label{Li14}
\end{figure}

Futhermore, orbital splitting effect can be generalized to explain the stabilities of alkali metal systems with ellipsoidal structures, such as the Li$_{14}$ cluster.  In the literature \cite{cheng_2013_communication}, the stability of the Li$_{14}$ cluster is explained as the combination of two Li$_{10}$ clusters with super-covalent bonds, mimicing fluorine molecule. However, superatomic orbitals of Li$_{14}$ from 43(LUMO)-48 do not correspond to any orbital of fluorine molecule according to our molecular orbital analysis. Within the view of orbital splitting effect, the structure and stability can be successfully interpreted. As illustrated in FIG. \ref{Li14}, the prolate structure of Li$_{14}$ will lead to the split of 1D energy levels. As a result, $1\mathrm{D}_{z^2}$ lies at the lowest energy levels, following the next higher energy of $1\mathrm{D}_{xz/yz}$ MOs. The $1\mathrm{D}_{z^2}$ and $1\mathrm{D}_{xz}/1\mathrm{D}_{yz}$ are all doubly occupied resulting in a large HOMO-LUMO gap of 1.60 eV.  This interpretation based on splitted orbitals clearly shows the nonspherical clusters can achieve good stability without meeting the requirement of magic number.\\
\subsection{Prediction of Geometric Structure}
\begin{figure}[H]
\centering
\includegraphics[scale=0.5]{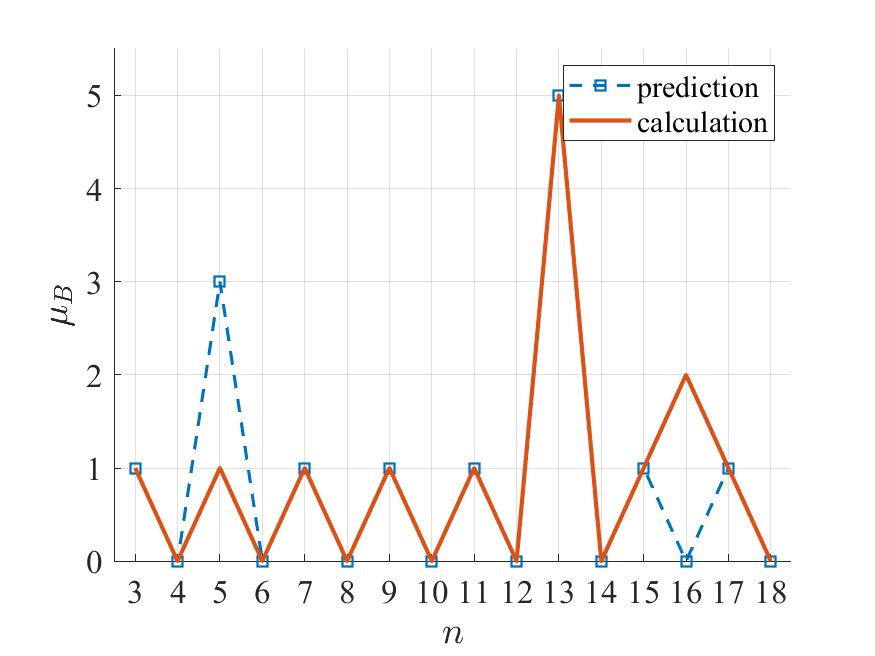}
\caption{Predicted magnetic moments for Li$_n$ clusters compared with the results calculated by DFT.}
\label{mub}
\end{figure}

It should be pointed out that the orbital splitting effect of Li$_{19}$ cluster proposed here can be extended to describe and predict the geometric shapes of the pure alkali metal species with other size. Based on previous discussion, the electron configuration of oblate, prolate and spherical jellium clusters can be theoretically predicted, and we are able to fill electrons in properly to connect the shape of cluster with the number of valence electrons ($n$). The rule of filling electrons are proposed in analogy with single element atoms: 
\begin{enumerate}
\item In all possible electron configurations, the term with fully occupied (degenerate) HOMOs has the lowest energy. This rule is based on the experimental fact that high peaks in mass spectrum correspond to closed superatom shells \cite{knight_1984_electronic, knight_electronic_1985}. 
\item When a closed shell cannot be achieved, the configuration with halfly occupied (degenerate) HOMOs has the lowest energy.  
\item When previous goals can't be satisfied, considering the fact that repulsion of electrons may raise the energy, we suppose HOMOs with less electrons occupied has lower energy. 
\end{enumerate}
\begin{figure*}[htbp]
\centering
\includegraphics[scale=0.1]{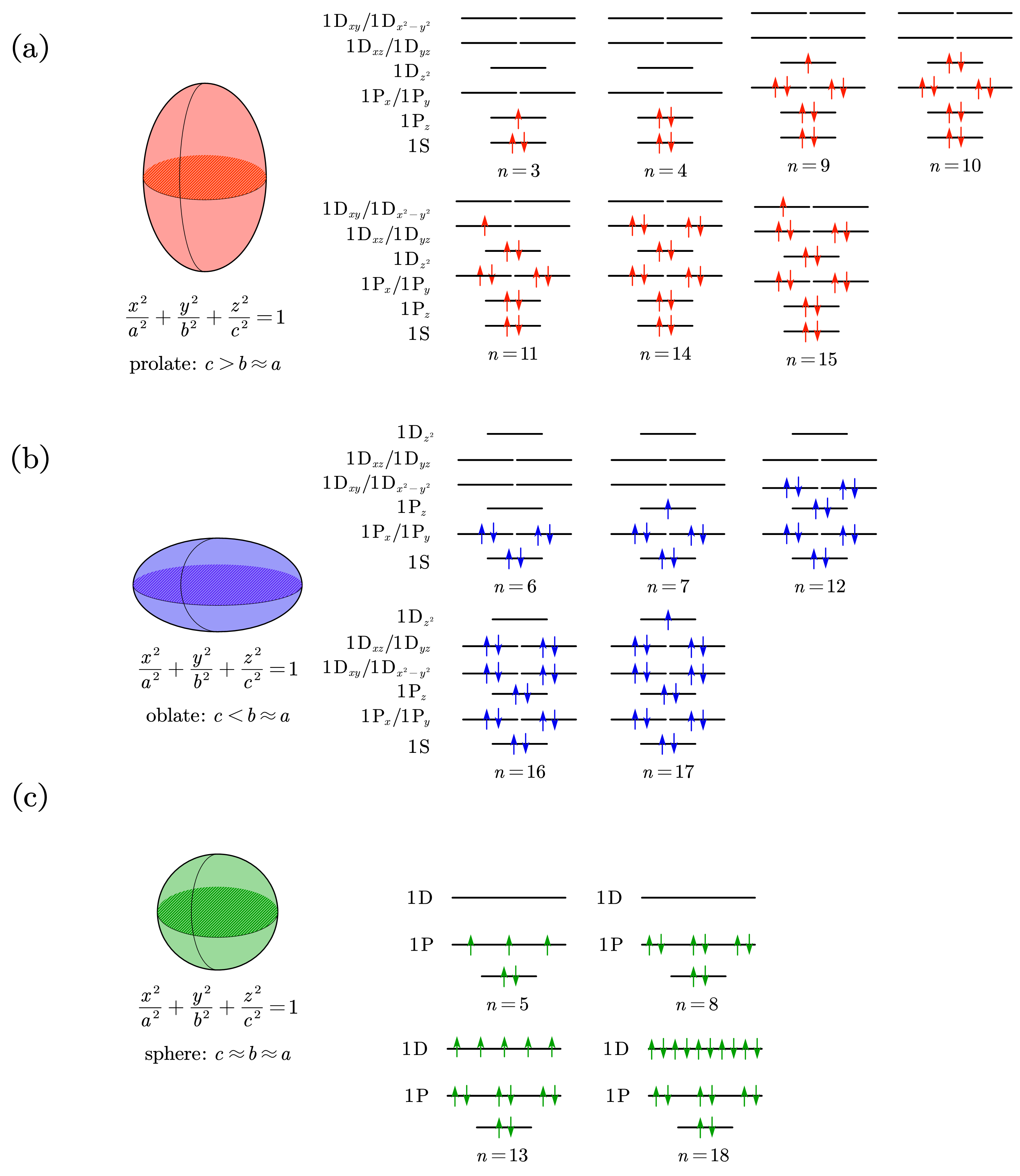}
\caption{Preferred electron occupation for (a) prolate, (b) oblate and (c) spherical clusters.}
\label{occupy}
\end{figure*}
As shown in FIG. \ref{occupy}, when $n=3(4)$, electrons can halfly (fullly) fill $1\mathrm{P}_z$ orbital, so the shape of cluster should be prolate. When $n=5$, three electrons halfly occupy three degenerate 1P orbitals, so the cluster will be shperical. When $n=6$, electrons can either halfly occupy $1\mathrm{P}_{x/y}$ orbitals with fully occupied $1\mathrm{P}_z$ in prolate configuration or fully occupy $1\mathrm{P}_{x/y}$ orbitals with empty $1\mathrm{P}_z$ in oblate configuration. Since fully occupied HOMOs are more stable, the cluster will be prolate. When $n=11$, the cluster will be prolate because in this case there will be only one electron in the HOMOs instead of three when the cluster is oblate. To sum up, for prolate clusters, numbers match electron configuration are $n = 3, 4, 9, 10, 11, 14, 15, \cdots$ For a cluster with oblate shape, the occupancy agrees with $n = 6, 7, 12, 16, 17, \cdots$ For spherical clusters, preferred numbers are $n=5, 8, 13, 18 \cdots$ The results are summarized in TABLE.\ref{shape-pref}. \\
\begin{table}[htbp]
  \centering
  \caption{Preferred shapes for clusters with different number of valence electrons($n$). '$\surd$' means 'preferred'.}
  \begin{ruledtabular}
    \begin{tabular}{cccccccc}
    n    & prolate & oblate & spherical & n    & prolate & oblate & spherical \\
    \colrule
    3    & $\surd$   &      &      & 11   & $\surd$   &      &  \\
    4    & $\surd$   &      &      & 12   &      & $\surd$   &  \\
    5    &      &      & $\surd$   & 13   &      &      & $\surd$ \\
    6    &      & $\surd$   &      & 14   & $\surd$   &      &  \\
    7    &      & $\surd$   &      & 15   & $\surd$   &      &  \\
    8    &      &      & $\surd$   & 16   &      & $\surd$   &  \\
    9    & $\surd$   &      &      & 17   &      & $\surd$   &  \\
    10   & $\surd$   &      &      & 18   &      &      & $\surd$ \\
    \end{tabular}%
    \end{ruledtabular}
  \label{shape-pref}%
\end{table}%
Once the configuration is determined, the magnetic moment of each cluster equal to the number of unpaired electrons (FIG. \ref{mub}). Compared with DFT calculated results, the prediction reproduced odd-even oscillation of magnet moments. In particular, the predicted magnetic moment of Li$_{13}$ clusters is 5$\mu_B$, in good agreement with the DFT theory calculations. It seems that we have overestimated the magnet moment in Li$_{5}$ while underestimated that in Li$_{16}$, the reasons will be analysed later. \\
\begin{figure*}[htbp]
\centering
\includegraphics[scale=0.1]{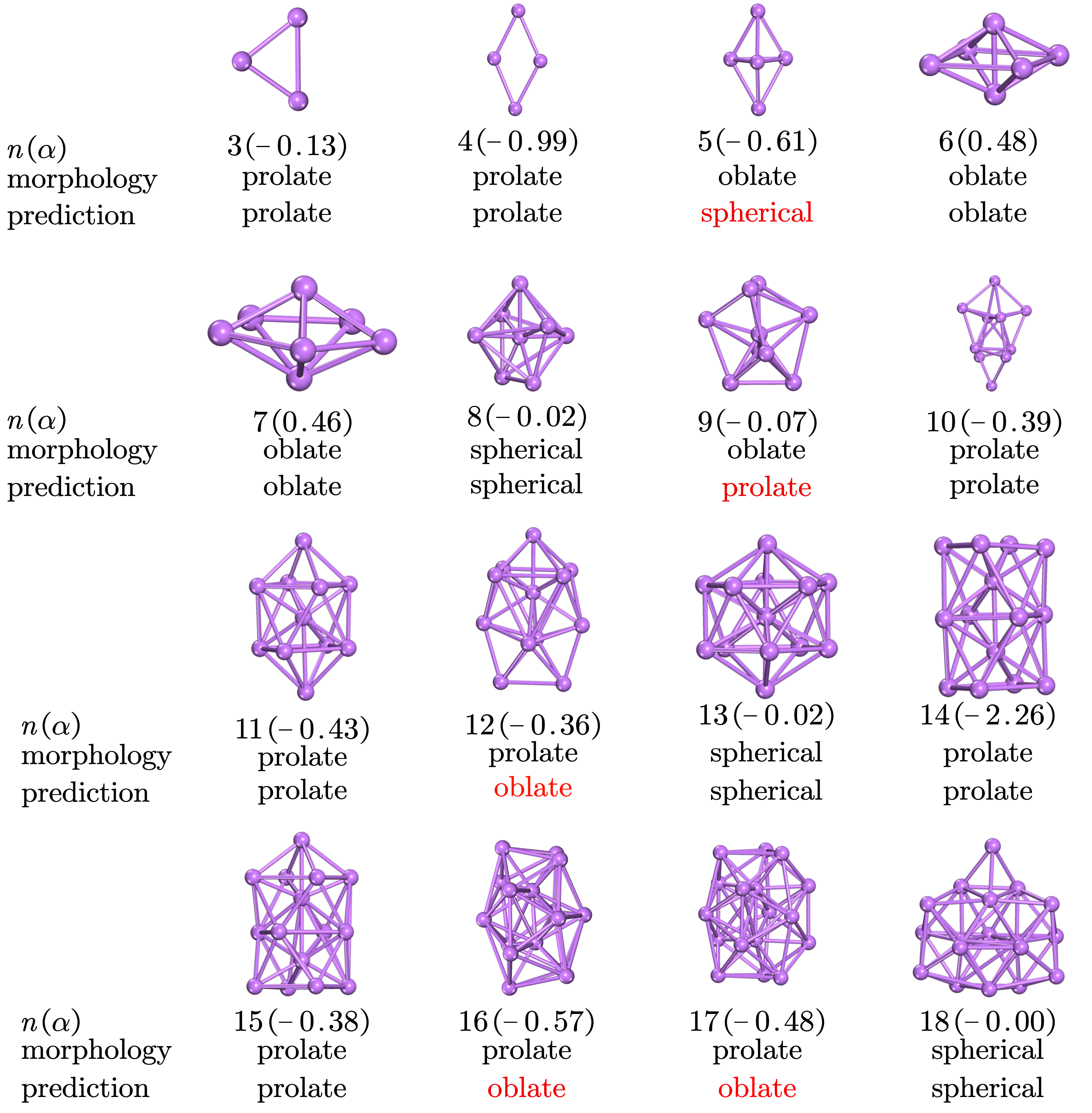}
\caption{Lowest energy configurations of Li$_n(n=3-18)$ clusters \cite{zhang_magnetic_2013, fournier_theoretical_2003}. Fitted $\alpha$ is applied to specify the mophorlogy (prolate, oblate or shperical) of the cluster. }
\label{pred}
\end{figure*}
When considering the lowest energy structures of $\mathrm{Li}_n$ clusters (FIG. \ref{pred}), in line with expectations, most of the shapes satisfy our prediction. When $n = 3$, instead of forming an equilateral triangle, Li$_3$ forms an isosceles triangle which can be viewed as a prolate structure stretched from a equilateral triangle in order to make the cluster more stable. Li$_4$ ($n=4$) has a tendency to be prolate so it will not form a regular tetrahedral (a square) which is close to spherical (oblate). For Li$_{10}$, Li$_{11}$, Li$_{14}$ and Li$_{15}$, the optimized equilibrium structures are prolate ellipsoid with their orbital configurations matching FIG. \ref{occupy} perfectly. \\
The most stable structures of Li$_{6}$ and Li$_{7}$ are all oblate ellipsoids, namely, Li$_{6}$ and Li$_{7}$ are squared pyramidal and pentagonal bipyramid respectively, which are consistent with theoretical prediction.\\
Li$_{8}$, Li$_{13}$ and Li$_{18}$ clusters have quite spherical shape and with high $\mathrm{D}_{5\mathrm{h}}$,  $\mathrm{I}_{\mathrm{h}}$, $\mathrm{C}_{\mathrm{s}}$ symmetry respectively. Li$_{13}$ has 5 parallel 1D orbitals resulting in a high magnetic moment of 5$\mu_B$, as forecasted in FIG. \ref{mub}. We can see that the model of orbital splitting can be successfully applied to interpret the structure and stability of other alkali metal clusters and the model is likely to be extended to $n>18$ clusters after taking into accunt of 1F, 2S, 2P... orbitals. It is also expected to be useful for the study of other types of doped metal clusters.\\
Although the geometric structures of the pure lithium clusters is overwhelmingly consistent with our predictions, there are also a few deviations. These exceptions may arise due to the multiple factors that determine the cluster structure. First of all, Jahn-Teller distortion is known to reduce the symmetry of small clusters and cause a low spin state \cite{khanna_1987_jahn}. Li$_{5}$ was set to be a high spin sphere with $3\mu_B$ but transformed to be a prolate spheroid with $1\mu_B$. Li$_{9}$ was predicted to be prolate but turned out spherical with no explicit $z$ direction. Likewise, Li$_{12}$ transformed from a prolate sphroid to a tri-axial ellipsoid with $a<b<c$ which leads to nondegeneration of the $1\mathrm{D}_{xz}$ and $1\mathrm{D}_{yz}$. The final electron configuration for Li$_{12}$ is $1\mathrm{S}^2-1\mathrm{P}_{z}^{2}-1\mathrm{P}_{y}^{2}-1\mathrm{P}_{x}^{2}-1\mathrm{D}_{z^2}^{2}-1\mathrm{D}_{yz}^{2}\left( \mathrm{HOMO} \right) -1\mathrm{D}_{xz}\left( \mathrm{LOMO} \right) $. Moreover, Hund's rule is reported to favour high spin state in cluster \cite{medel_hunds_2011, geguzin_1981_spin}, which may be responsible for prolate Li$_{16}$  with $2\mu_B$.  In addition, we ignored the intercalation of 2S orbitals. It has been calculated that 2S inserted between $1\mathrm{D}_{xy/x^2-y^2}^{4}$ and $1\mathrm{D}_{xz/yz}^{4}$ in Li$_{19}$ while it inserted between 1P and 1D in Li$_{6}$. Li$_{17}$ is a prolate ellipsoid which has 2S orbitals intercalated in $1\mathrm{D}_{xy/x^2-y^2}^{}$, causing a orbital configuration of $1\mathrm{S}^2-1\mathrm{P}_{z}^{2}-1\mathrm{P}_{x/y}^{4}-1\mathrm{D}_{z^2}^{2}-1\mathrm{D}_{xz/yz}^{4}-1\mathrm{D}_{xy/x^2-y^2}^{3}\left( \mathrm{HOMO} \right) -2\mathrm{S}\left( \mathrm{LOMO} \right)-1\mathrm{D}_{xy/x^2-y^2}^{} $. There only exsist a small gap (0.148 eV) between HOMO and LUMO. It is predictable that this effect will be negligible for $n>20$ clusters.\\
\section{CONCLUSION}
In conclusion, our work offered a method to predict the magnetic moment and shape of alkali clusters through the analysis of superatom orbitals. The molecular orbitals of nine low-lying Li$_{19}$ clusters have been calculated, identified and compared. It has been found that clusters in the shape of oblate ellipsoid have the electronic configuration of $1\mathrm{S}^2-1\mathrm{P}^6-1\mathrm{D}_{xy/x^2-y^2}^4-2\mathrm{S}^2-1\mathrm{D}_{xz/yz}^4-1\mathrm{D}_{z^2}^1$, while clusters in the shape of prolate ellipsoid have the sequence of  $1\mathrm{S}^2-1\mathrm{P}^6-1\mathrm{D}_{z^2}^2-1\mathrm{D}_{xz/yz}^4-1\mathrm{D}_{xy/x^2-y^2}^4-2\mathrm{S}^1$. For originally degenerate orbitals, the more the orbital extends in the direction of the cluster extension, the lower the energy becomes. The applications of the orbital splitting are promoted in two following ways. The stability of the Li$_{14}$ cluster are explained in the framework of orbital splitting effect without resorting to the SVB model.  Meanwhile, it is proposed that the orbital splitting effect can be used to predict the shape and magnetic moment of clusters.  11 out of 16 predicted shapes of Li$_n(n=3-18)$ are consistent with the results obtained by the principle of minimum energy.
\begin{acknowledgements}
The authors thank ShangHai Students' Innovation and Entrepreneurship Program (No.S20091) for supporting this research. 
\end{acknowledgements}
\bibliographystyle{unsrt} 
\bibliography{reference}

\begin{thebibliography}{10}

\bibitem{ekardt_work_1984}
W.~Ekardt.
\newblock Work function of small metal particles: {Self}-consistent spherical
  jellium-background model.
\newblock {\em Physical Review B}, 29(4):1558--1564, February 1984.

\bibitem{ekardt_theory_1986}
W.~Ekardt.
\newblock Theory of electronic excitations in coated metal particles:
  {Jellium}-on-jellium model.
\newblock {\em Physical Review B}, 34(2):526--533, July 1986.

\bibitem{shao_density_2015}
Peng Shao, Li-Ping Ding, Hai-Tao Feng, and Jiang-Tao Cai.
\newblock Density function study transition metal chromium-doped alkali
  clusters: the finding of magnetic superatom.
\newblock {\em Molecular Physics}, 113(11):1337--1346, June 2015.

\bibitem{yan_face-sharing_2019}
Lijuan Yan.
\newblock Face-{Sharing} {Homo}- and {Hetero}-{Bitetrahexahedral} {Superatomic}
  {Molecules} {M} $_{\textrm{1}}$ {M} $_{\textrm{2}}$ @{Li} $_{\textrm{20}}$
  ({M} $_{\textrm{1}}$ /{M} $_{\textrm{2}}$ = {Ti} and {W}).
\newblock {\em The Journal of Physical Chemistry A}, 123(26):5517--5524, July
  2019.

\bibitem{yan_superatomic_2020}
Lijuan Yan, Jun Liu, and Jianmei Shao.
\newblock Superatomic properties of transition-metal-doped tetrahexahedral
  lithium clusters: {TM}@{Li} $_{\textrm{14}}$.
\newblock {\em Molecular Physics}, 118(2):e1592256, January 2020.

\bibitem{sung_packing_1994}
Ming-Wen Sung, Ryoichi Kawai, and John~H. Weare.
\newblock Packing {Transitions} in {Nanosized} {Li} {Clusters}.
\newblock {\em Physical Review Letters}, 73(26):3552--3555, December 1994.

\bibitem{zhang_magnetic_2013}
Meng Zhang, Jianfei Zhang, Xiaojuan Feng, Hongyu Zhang, Lixia Zhao, Youhua Luo,
  and Wei Cao.
\newblock Magnetic {Superatoms} in {VLi} $_{\textrm{ \textit{n} }}$ (
  \textit{n} = 1–13) {Clusters}: {A} {First}-{Principles} {Prediction}.
\newblock {\em The Journal of Physical Chemistry A}, 117(48):13025--13036,
  December 2013.

\bibitem{zhang_probing_2010}
Meng Zhang, Xiao-Yan Gu, Wen-Li Zhang, Li-Na Zhao, Li-Ming He, and You-Hua Luo.
\newblock Probing the magnetic and structural properties of the 3d, 4d, 5d
  impurities encapsulated in an icosahedral {Ag12} cage.
\newblock {\em Physica B: Condensed Matter}, 405(2):642--648, January 2010.

\bibitem{pauna_evolution_2017}
Henri Pauna, Xinying Shi, Marko Huttula, Esko Kokkonen, Taohai Li, Youhua Luo,
  Jyrki Lappalainen, Meng Zhang, and Wei Cao.
\newblock Evolution of lithium clusters to superatomic {Li} $_{\textrm{3}}$ {O}
  $^{\textrm{+}}$.
\newblock {\em Applied Physics Letters}, 111(10):103901, September 2017.

\bibitem{cheng_2014_superatom}
Longjiu Cheng, Xiuzhen Zhang, Baokang Jin, and Jinlong Yang.
\newblock Superatom--atom super-bonding in metallic clusters: a new look to the
  mystery of an au 20 pyramid.
\newblock {\em Nanoscale}, 6(21):12440--12444, 2014.

\bibitem{li_au20_2003}
J.~Li.
\newblock Au20: {A} {Tetrahedral} {Cluster}.
\newblock {\em Science}, 299(5608):864--867, February 2003.

\bibitem{bergeron_formation_2004}
D.~E. Bergeron.
\newblock Formation of {Al13I}-: {Evidence} for the {Superhalogen} {Character}
  of {Al13}.
\newblock {\em Science}, 304(5667):84--87, April 2004.

\bibitem{reveles_2009_designer}
J.~U. Reveles, P.~A. Clayborne, A.~C. Reber, S.~N. Khanna, K.~Pradhan, P.~Sen,
  and M.~R. Pederson.
\newblock Designer magnetic superatoms.
\newblock {\em Nature Chemistry}, 1(4):310, 2009.

\bibitem{kugel_1982_jahn}
Kliment~I Kugel and DI~Khomskii.
\newblock The jahn-teller effect and magnetism: transition metal compounds.
\newblock {\em Soviet Physics Uspekhi}, 25(4):231, 1982.

\bibitem{cheng_2013_communication}
Longjiu Cheng and Jinlong Yang.
\newblock Communication: New insight into electronic shells of metal clusters:
  Analogues of simple molecules, 2013.

\bibitem{yan_seventeen-coordinate_2019}
Lijuan Yan, Jianmei Shao, Liren Liu, and Chunlei Chen.
\newblock Seventeen-coordinate binary metal superatoms: {M}@{Li17}.
\newblock {\em Chemical Physics Letters}, 733:136693, October 2019.

\bibitem{lv_2012_particle}
Jian Lv, Yanchao Wang, Li~Zhu, and Yanming Ma.
\newblock Particle-swarm structure prediction on clusters.
\newblock {\em The Journal of chemical physics}, 137(8):084104, 2012.

\bibitem{wang_1991_correlation}
Yue Wang and John~P Perdew.
\newblock Correlation hole of the spin-polarized electron gas, with exact
  small-wave-vector and high-density scaling.
\newblock {\em Physical Review B}, 44(24):13298, 1991.

\bibitem{luo_special_2014}
Zhixun Luo and A.~Welford Castleman.
\newblock Special and {General} {Superatoms}.
\newblock {\em Accounts of Chemical Research}, 47(10):2931--2940, October 2014.

\bibitem{clemenger_ellipsoidal_1985}
Keith Clemenger.
\newblock Ellipsoidal shell structure in free-electron metal clusters.
\newblock {\em Physical Review B}, 32(2):1359--1362, July 1985.

\bibitem{knight_1984_electronic}
WD~Knight, Keith Clemenger, Walt~A de~Heer, Winston~A Saunders, MY~Chou, and
  Marvin~L Cohen.
\newblock Electronic shell structure and abundances of sodium clusters.
\newblock {\em Physical review letters}, 52(24):2141, 1984.

\bibitem{knight_electronic_1985}
W.D. Knight, Walt~A. de~Heer, Keith Clemenger, and Winston~A. Saunders.
\newblock Electronic shell structure in potassium clusters.
\newblock {\em Solid State Communications}, 53(5):445--446, February 1985.

\bibitem{fournier_theoretical_2003}
René Fournier, Joey Bo~Yi~Cheng, and Anna Wong.
\newblock Theoretical study of the structure of lithium clusters.
\newblock {\em The Journal of Chemical Physics}, 119(18):9444--9454, November
  2003.

\bibitem{khanna_1987_jahn}
SN~Khanna, BK~Rao, P~Jena, and JL~Martins.
\newblock Jahn-teller distortion, hund’s coupling and metastability in alkali
  tetramers.
\newblock In {\em Physics and Chemistry of Small Clusters}, pages 435--438.
  Springer, 1987.

\bibitem{medel_hunds_2011}
V.~M. Medel, J.~U. Reveles, S.~N. Khanna, V.~Chauhan, P.~Sen, and A.~W.
  Castleman.
\newblock Hund's rule in superatoms with transition metal impurities.
\newblock {\em Proceedings of the National Academy of Sciences},
  108(25):10062--10066, June 2011.

\bibitem{geguzin_1981_spin}
II~Geguzin.
\newblock Spin polarization of high-symmetry clusters of simple metals in the
  ground state.
\newblock {\em ZhETF Pisma Redaktsiiu}, 33:584, 1981.

\end{thebibliography}
\end{document}